\documentclass[smallextended]{svjour3}       
\smartqed  
\usepackage{graphicx}
\begin{document}

\title{Atom-field entanglement in  two-atom Jaynes-Cummings model with intensity-dependent coupling
}

\titlerunning{Atom-field entanglement in  two-atom Jaynes-Cummings model }        

\author{E.K. Bashkirov
}


\institute{E.K. Bashkirov          \at
              Department of General and Theoretical Physics, Samara State University, 1 Academican Pavlov St., Samara,  443011,  Russia \\
              Tel.: +123-45-678910\\
              Fax: +123-45-678910\\
              \email{bash@samsu.ru}           
}

\date{Received: date / Accepted: date}

\maketitle

\begin{abstract}
An exact solution of the problem of  two-atom  one- and two-mode Jaynes-Cummings model with intensity-dependent coupling is presented.
 Asymptotic solutions for system state vectors are obtained in the approximation of large initial coherent fields.
 The atom-field entanglement is investigated on the basis of the reduced
atomic entropy dynamics. The possibility of the system being initially in a pure disentangled
state to revive into this state during the evolution process for both models is shown. Conditions and  times
 of disentanglement are derived.

\end{abstract}

\section{Introduction}
\label{intro}

Entanglement  plays a central role in quantum information, quantum computation and communication, and
quantum cryptography.  In recent years, there has
been a considerable effort to characterize entanglement properties qualitatively and
quantitatively   and to apply them in
quantum information. A lot of schemes are proposed for
many-particle entanglement generation. The simplest scheme to
investigate the atom-field entanglement is the
 Jaynes-Cummings model (JCM) \cite{jc} describing an interaction of a
two-level atom with a single-mode quantized radiation field. This
model is of fundamental importance for quantum optics
\cite{Yoo,Shore} and is realizable to a very good approximation
in experiments with Rydberg atoms in high-Q superconducting
cavities, trapped ions, superconducting circuits etc. \cite{Walter,Haroche,Nori}. The model predicts a variety of
interesting phenomena. The atom-field entanglement  is among
them. An investigation of the atom-field entanglement for JCM has been
initiated by Phoenix and
Knight \cite{Knightearly},\cite{Knight} and Gea-Banacloche
\cite{Gea-Banacloche},\cite{Gea-Banacloche1}. Gea-Banacloche has derived an asymptotic
result  for the JCM state vector which is valid when the field is
initially in a coherent state with a large mean photon number.
It is shown that the atom prepared in arbitrary initial pure
atomic state is to a good approximation in a pure state in the
middle of the collapse region. This has been first noticed by
Phoenix and Knight  by using the entropy concepts.  An
appreciable disentanglement between atom and field is found at
the half-revival time, otherwise the atom and field are strongly
entangled. Moreover, at the half-revival time, the cavity field
represents a coherent superposition of the two macroscopically
distinct states with opposite phases or so-called Schr$\ddot{\rm
o}$edinger cat state.   The
theory outlined in \cite{Gea-Banacloche} has been generalized
for  two-photon JCM \cite{Dung1},\cite{Nasreen}, two-photon JCM
with nondegenerate two-photon and Raman transitions
\cite{Abdalla},\cite{Larson}, two-atom JCM \cite{Kudryavtsev}, \cite{Dung},  two-atom one-mode Raman
coupled model \cite{Rdegen} and two-atom two-photon JCM \cite{bash1}-\cite{bash3}.

Two-photon processes are known to play a very important role in
atomic systems due to high degree of correlation
 between  emitted photons. An interest for investigation of the two-photon JCM is stimulated by
the experimental realization of a two-photon one-atom
micromaser on Rydberg transitions in a microwave cavity
\cite{Brune}. A nondegenerate two-photon two-mode maser, which
represents a two-level Rydberg atom interacting with two
different modes of a quantum electromagnetic field in a
high-quality cavity through a nondegenerate two-photon transition, is
an important generalization of the model of a two-photon
micromaser. A possibility of the modulation, amplification and
control of one mode with another mode is an important feature of
the two-photon two-mode maser. JCM with nondegenerate two-photon
transitions  have attracted a great
deal of attention. The foregoing model  have been
considered in  terms  of atomic population dynamics research,
field statistics research, field and atom squeezing analysis,
atom and field entropy and entanglement examining \cite{Del}.
The  two-atom two-photon JCM for initial two-mode coherent cavity
field has been investigated for   nondegenerate  two-photon transitions in \cite{Bash}.

As it was pointed out by Singh and Amrita \cite{Singh}  the Cavity Quantum Electrodynamics (QED) generally
deals with few cavity photons, hence, atomic emission and absorption effects are expected to change the
atom-field interaction strength significantly. Consequently
an intensity-dependent coupling constant  would
be appropriate to study the problems related to cavity
QED. The dynamical properties of the intensity-dependent one- and two-atom
JCM for a two-mode cavity field have been investigated recently in Refs \cite{Singh}-\cite{Hekmatara}.

In this paper we analyze the atomic  and field state evolution
and atom-field entanglement in the  two-atom one- and two-mode JCM with  intensity-dependent coupling, as in
\cite{Gea-Banacloche1}, that the field is initially in a one- or two-mode
coherent state with a large  mean photon numbers correspondingly. We study
systems by using the linear atomic entropy concepts and the asymptotic behavior  of the system state vectors  in the approximation of large initial coherent field. The
main goal of this paper is to show  such initial states of
atomic subsystem which provide disentanglement between atom and
field at certain times. We also have estimated these disentanglement
times. In the framework of large initial field the atomic eigenstates of the semiclassical
Hamiltonian are found. It is shown that  if  atoms are
initially prepared in one of these states, the system evolves
remaining the atomic and field parts separately disentangled in a
pure state. However, only for certain initial atomic states  the
disentanglement between atom and field occurs.

\section{Model description. The exact solution of Schr$\ddot{\rm
o}$edinger equation for wave function}
\label{sec:1}

We have investigated atom-field entanglement for
two typeû of  two-atom JCM  with the intensity-dependent coupling.
The first of them  describes two two-level atoms  resonantly
interacting with  one-mode coherent  field   in lossless cavity.  The  interaction Hamiltonian of
such a model is
$$
H_{int}=\hbar g\sum_{i=1}^{2} \left(\sqrt{a^+ a} a^+ \sigma_i^-  + \sigma_i^+ a \sqrt{a^+ a} \right),
\eqno{(1)}
$$
Here  we use the following notationû: $a^+\,(a)$ is creation (annihilation) operator for cavity mode, $\sigma_i^+ = |+\rangle_{ii}\langle -| $ and $\sigma_i^-=|-\rangle_{ii}\langle + |$ are the atomic transition operators, while $|-\rangle_i$ and $ |+\rangle_i$ denote the ground and excited states of the
$i$th  two-level atom ($i=1,\,2$) respectively. Parameter  $g$ with the operator $\sqrt{a^+ a}$ plays the role of intensity-dependent coupling constant between atoms and the cavity field.
The second model  describes two
two-level atoms  interacting with two-mode coherent field in lossless cavity via nondegenerate two-photon transitions under the assumption of exact
two-photon resonance. The
effective interaction Hamiltonian for considered model can be
written in the following form \cite{Singh2}:
$$
H_{int}=\hbar
g\sum^2_{i=1}{(\sqrt{a^+_1a_1}a^+_1\sqrt{a^+_2a_2}a^+_2\sigma^-_i+a_1\sqrt{a^+_1a_1}{a_2\sqrt{a^+_2a_2}\sigma}^+_i)}. \eqno{(2)}
$$
In formulae (2) we use the following notations: $a_{j}^+\,(a_{j})$ is creation (annihilation) operator
for $j$-th cavity mode ($j=1,2)$ and $g$ is the  effective constant of dipole-photon interaction. In two-photon processes the Stark shift caused by the
intermediate atomic level plays the role of an intensity-dependent detuning. However,  if the two fields are
tuned in such a way that both have reverse detuning with
the intermediate atomic level, the Stark shift will not appear as has been pointed in Ref. \cite{Singh2}. Such a two-photon signal could be achieved by two
dye lasers.

Atoms are supposed to be initially prepared in arbitrary pure atomic states superposition
$$|\Psi(0)\rangle_A =\alpha |+,+\rangle + \beta |+,-\rangle + \gamma |-,+\rangle + \delta |-,-\rangle $$
and field is supposed to be initially in one-mode or two-mode coherent state. The full wave function of atom-field system at initial time for the first model can be written as
$$|\Psi(0)\rangle =|\Psi(0)\rangle_A  |\upsilon\rangle$$
and for the second model can be written as
$$|\Psi(0)\rangle =|\Psi(0)\rangle_A  |\upsilon_{1}\upsilon_{2} \rangle.$$
Here $\alpha, \beta, \gamma $ and $\delta$ are
 arbitrary complex values satisfying  the  condition
$$ |\alpha|^2 + |\beta|^2 + |\gamma|^2 + |\delta|^2 = 1.$$
Here
$$|\upsilon \rangle =\sum_{n=0}^{\infty} F_{n}|n\rangle
$$
is one-mode and
$$|\upsilon_{1}\upsilon_{2} \rangle =\sum_{n_1=0}^{\infty}\sum_{n_2=0}^{\infty} F_{n_1}F_{n_2}|n_1\rangle |n_2\rangle \>
$$
is two-mode coherent state correspondingly, where  the coefficients $F_{n}$ are
$$F_{n}=exp(-\overline{n} /2)\frac{\overline{n}^{n/2}}{\sqrt{n!}}e^{\imath  \varphi}\,\>,
$$
where $ \upsilon = \overline{n}^{1/2}e^{\imath \varphi}$, $
\overline{n} = |\upsilon|^2 $ is the mean photon number and
$\varphi$ is the phase of  coherent mode.

The exact
solution of Shr$\ddot{\rm o}$edinger equation for wave function
under considered initial conditions  for model with
Hamiltonian (1)  takes the form of:
$$
|\Psi(t)\rangle=\sum_{n}\left[ A_{n}(t)|+,+\rangle +  B_{n}(t)|+,-\rangle +  C_{n}(t)|-,+\rangle + D_{n}(t)|-,- \rangle \right ]|n\rangle. \eqno{(3)}
$$
Here for one-mode two-atom JCM the following notation is accepted:
\begin{eqnarray*}
A_{n}(t)&=& \frac{(n+2)^2 + (n+1)^2 \cos(2 \Omega_n t)}{2 \Omega_n^2} \alpha F_n  -
\imath \frac{(n+1) \sin(2 \Omega_n )}{2 \Omega_n } \beta F_{n+1} - \nonumber\\
&-&\imath \frac{(n+1) \sin(2 \Omega_n t)}{2 \Omega_n } \gamma F_{n+1} - \frac{(n+1)(n+2)\sin^2(\Omega_n t)}{\Omega_n^2}  \delta F_{n+2},\nonumber\\
B_{n}(t)&=& -\imath \frac{n\sin(2 \Omega_{n-1} t)}{2\Omega_{n-1}} \alpha F_{n-1}  +
 \cos^2( \Omega_{n-1} t) \beta F_{n} -\nonumber\\
&-&  \sin^2( \Omega_{n-1} t) \gamma F_{n} - \imath \frac{(n+1)\sin^2(\Omega_{n-1} t)}{2\Omega_{n-1}}  \delta F_{n+1},\nonumber\\
C_{n}(t)&=& -\imath \frac{n\sin(2 \Omega_{n-1} t)}{2\Omega_{n-1}} \alpha F_{n-1} -  \sin^2( \Omega_{n-1} t) \beta F_{n}  + \nonumber\\
&+&
 \cos^2(\Omega_{n-1} t) \gamma F_{n} - \imath \frac{(n+1)\sin^2(\Omega_{n-1} t)}{2\Omega_{n-1}}  \delta F_{n+1},\nonumber\\
D_{n}(t)&=& - \frac{(n-1)n\sin^2( \Omega_{n-2} t)}{\Omega_{n-2}^2}  \alpha F_{n-2}  - \imath \frac{n \sin(2 \Omega_{n-2} t)}{2 \Omega_{n-2} } \beta F_{n-1} - \nonumber\\
&-& \imath \frac{n \sin(2 \Omega_{n-2} t)}{2 \Omega_{n-2} } \gamma F_{n-1} + \frac{((n-1)^2 + n^2) \cos^2(\Omega_{n-2} t)}{2\Omega_{n-2}^2}  \delta F_{n},\nonumber\\
\end{eqnarray*}
where
$$\Omega_n=\sqrt{\left[2n (n+3)+5 \right ]/2}.$$

The exact solution of Shr$\ddot{\rm o}$edinger
equation for wave function under considered initial conditions
for model with Hamiltonian (2) is
$$
|\Psi(t)\rangle =\sum^{\infty}_{n_1,n_2=0}\left[A_{n_1n_2}(t) |+,+\rangle
+ B_{n_1n_2}t)|+,- \rangle + \right.$$ $$\left.
+C_{n_1n_2}(t)|-,+\rangle +D_{n_1n_2}(t)|-,-\rangle
\right) |n_1,n_2\rangle. \eqno{(4)}
$$
Here the following notation is accepted:
\begin{eqnarray*}
A_{n_1n_2}\left(t\right)&=&\frac{2\alpha F_{n_1}F_{n_2}}{{\Omega }^2_1}\left(X_1\left(n_1+1\right)\left(n_2+1\right){\cos  {\Omega }_{{\rm 1}}{\rm t+}{{\rm X}}_{{\rm 2}}{(n}_1+2)(n_2+2)\ }\right)+\nonumber\\
&+&\frac{2\beta F_{n_1+2}F_{n_2+2}}{{\Omega }^2_1}X^2_2\left(cos{\Omega }_{{\rm 1}}t-1\right)- +\nonumber\\
&-&i\frac{\left(\gamma +\delta \right)F_{n_1+1}F_{n_2+1}}{{\Omega }_{{\rm 1}}}\left(n_1+1\right)\left(n_2+1\right)sin{\Omega }_{{\rm 1}}{\rm t,}\nonumber\\
B_{n_1n_2}\left(t\right)&=&-i\frac{\alpha F_{n_1-1}F_{n_2-1}}{{\Omega }_3}X_3sin{\Omega }_{{\rm 3}}t-i\frac{\beta F_{n_1+1}F_{n_2+1}}{{\Omega }_3}\left(n_1+1\right)\left(n_2+1\right)sin{\Omega }_{{\rm 3}}t+\nonumber\\
&+&\frac{1}{{\rm 2}}(\gamma -\delta +\left(\gamma +\delta \right)cos{\Omega }_{{\rm 3}}{\rm t)}F_{n_1}F_{n_2},\nonumber\\
C_{n_1n_2}\left(t\right)&=&-i\frac{\alpha F_{n_1-1}F_{n_2-1}}{{\Omega }_3}X_3sin{\Omega }_{{\rm 3}}t-i\frac{\beta F_{n_1+1}F_{n_2+1}}{{\Omega }_3}\left(n_1+1\right)\left(n_2+1\right)sin{\Omega }_{{\rm 3}}t+\nonumber\\
&+&\frac{1}{{\rm 2}}(\delta -\gamma +\left(\gamma +\delta \right)cos{\Omega }_{{\rm 3}}{\rm t)}F_{n_1}F_{n_2},\nonumber\\
D_{n_1n_2}\left(t\right)&=&\frac{2\beta F_{n_1}F_{n_2}}{{\Omega }^2_2}\left(X_4\left(n_1-1\right)\left(n_2-1\right)+X_3n_1n_2cos{\Omega }_{{\rm 2}}{\rm t}\right)+\nonumber\\
&+&\frac{2\alpha F_{n_1-2}F_{n_2-2}}{{\Omega }^2_2}X_3X_4\left(cos{\Omega }_{{\rm 2}}t-1\right)-+\nonumber\\
&-&i\frac{\left(\gamma +\delta \right)F_{n_1-1}F_{n_2-1}}{{\Omega }_{{\rm 2}}}X_3sin{\Omega }_{{\rm 2}}{\rm t,}\nonumber\\
\end{eqnarray*}
where
 $${\Omega }_{{\rm 1}}\left({{\rm n}}_{{\rm
1}},{{\rm n}}_{{\rm 2}}\right){\rm =}\sqrt{{\rm 2(}{{\rm X}}_{{\rm
1}}\left({{\rm n}}_{{\rm 1}}{\rm +1}\right)\left({{\rm n}}_{{\rm
2}}{\rm +1}\right){\rm +}{{\rm X}}_{{\rm 2}}{\rm (}{{\rm
n}}_{{\rm 1}}{\rm +2)(}{{\rm n}}_{{\rm 2}}{\rm +2))}}{\rm
=}$$ $${\rm=}{\Omega }_{{\rm 2}}\left({{\rm n}}_{{\rm 1}}{\rm +2,}{{\rm
n}}_{{\rm 2}}{\rm +2}\right){\rm =}{\Omega }_{{\rm 3}}\left({{\rm
n}}_{{\rm 1}}{\rm +1,}{{\rm n}}_{{\rm 2}}{\rm +1}\right)$$ are the Rabi frequencies, and
 $$X_1\left({{\rm n}}_{{\rm 1}},{{\rm
n}}_{{\rm 2}}\right)=\sqrt{{{\rm n}}_{{\rm 1}}{{\rm n}}_{{\rm
2}}\left({{\rm n}}_{{\rm 1}}{\rm +1}\right)\left({{\rm n}}_{{\rm
2}}{\rm +1}\right)}=$$ $$ = X_2\left({{\rm n}}_{{\rm 1}}-1,{{\rm
n}}_{{\rm 2}}-1\right)=X_3\left({{\rm n}}_{{\rm 1}}+1,{{\rm
n}}_{{\rm 2}}+1\right)=X_4({{\rm n}}_{{\rm 1}}+2,{{\rm n}}_{{\rm
2}}+2).
$$

By using the exact solution  (3) or (4) the reduced atomic density matrix can be constructed tracing the expression
$|\Psi(t)\rangle\langle\Psi(t)|$ over field variables. Thus, having the exact expression for the system wave function
we can obtain the accurate statement for atom-field entanglement parameter as well as evaluate the entanglement degree.

Using the state vector (3) or (4) one can calculate the  mean values of observable quantities.
For instance, for the first model the  probability to find both atoms
in the excited states is represented as
$$W_{++}(t) = \sum\limits_{n} |A_{n}(t)|^2. \eqno{(5)}$$
Probabilities  exhibit fast oscillations at frequencies $2\Omega_n$
and $4\Omega_n$. The interference of the
terms with different numbers of photons  should lead to the buildup and decay of Rabi
oscillations for probabilities  as in the case of
the conventional one-photon one-mode JCM
model. In contrast to the conventional
JCM, the expressions for the probabilities (5)
contain two types of fast-oscillating terms. Therefore, in the
case under study, two types of buildup and decay must
be realized for the Rabi oscillations of the atomic probabilities.
Let us estimate the revivals period of the Rabi
oscillations for large value of the mean photon number in cavity mode. Then, in the Taylor
expansion for the Rabi frequency in the vicinity
of $\overline{n}$, we take into account only terms of first
order with respect to deviations $ n  - \overline{n}$:
$$\Omega_n = \Omega_{\overline{n}} +\frac{2 \overline{n} +3}{\sqrt{2[2 \overline{n}(\overline{n}+3) +5]}}(n - \overline{n}).$$

For the terms in formula (5) that oscillate at frequencies $2\Omega_n$
and $4\Omega_n$  the recovery takes place at time intervals
$$|2\Omega_{\overline{n}+1} - 2\Omega_{\overline{n}}|T_{1R} = 2\pi k, \eqno{(6)}$$
$$|4\Omega_{\overline{n}+1} - 4\Omega_{\overline{n}}|T_{2R} = 2\pi m, \eqno{(7)}$$
where $k$ and $m = 0, 1, 2, …$.
For  high intensities of the coherent cavity
field mode ($\overline{n} \gg 1$), formulas (6) and (7) are represented
as $gT_{1R} = \pi k$ and  $gT_{2R} = \pi m/2$.
Thus, we have two series of the Rabi oscillation revivals for probabilities with periods $T_{1R}$ and $T_{2R}$. As have been shown in Ref. \cite{Singh2} for the second model  there are also two series of the Rabi oscillation revivals. For $\bar n_1, \bar n_2 \gg 1$ these periods are:
$$gT'_{1R} = \frac{\pi \sqrt{k}}{\sqrt{\bar n_1 \bar n_2}},\eqno{(8)}$$
$$gT'_{2R} = gT'_{1R}/2.\eqno{(9)}$$
Singh et al. \cite{Singh2} have derived by means of computer calculations that probabilities not always
show revivals at times predicted by Eqs. (8) and (9),
but when a revival occurs, it occurs at one of those times, and that in the time evolution
the probabilities  lost the periodic character, showing chaotic-like behavior.

\section{System state vector evolution}
\label{sec:2}

In Section I an exact solution for Shr$\ddot{\rm o}$edinger equation
(3) and (4) was obtained for the considered models with intensity-dependent coupling. By using this
solutions it is  possibile to obtain analytical results
for  atom-field entanglement. We will show that for the atoms and
field  prepared initially in  some pure disentangled states the
wave function at some moments of time could be factorized into
product of atomic and field subsystems state vectors. To obtain
this result we suppose that the field is initially in a coherent one-mode or two-mode state of high intensity
and examine the  time behavior of eigenvectors  of semiclassical interaction Hamiltonians.
By analyzing the asymptotic evolution of the mentioned vectors we can obtain the required initial
states.  The semiclassical
interaction Hamiltonians for two-atom one-mode and two-mode JCM with intensity-dependent coupling are
$$
H_{SC}=\hbar g |\upsilon| \sum\limits_{i=1}^2 \left(\upsilon^{*} \sigma_i^{-}+
\upsilon \sigma_{i}^{+}\right).
\eqno{(10)}
$$
and
$$H_{SC}=\hbar
g \sum\limits_{i=1}^2\left(\sqrt{\upsilon_{1}^{*}\upsilon_{1}}\upsilon_{1}^{*}\sqrt{\upsilon_{2}^{*}\upsilon_{2}}\upsilon_{2}^{*}\sigma_{i}^{-}+
\upsilon_{1}\sqrt{\upsilon_{1}^{*}\upsilon_{1}}\upsilon_{2}\sqrt{\upsilon_{2}^{*}\upsilon_{2}}\sigma_{i}^{+}\right).\eqno{(11)}$$
correspondingly.

The eigenvectors of  Hamiltonians (10) or (11)  are found to be
\begin{eqnarray}
|\Phi_1\rangle &=&\frac12\left[e^{2i\theta} |+,+\rangle +|-,-\rangle +e^{i\theta}\left(|+,-\rangle
+|-,+\rangle \right)\right],\nonumber\\
|\Phi_2\rangle &=&\frac12\left[e^{2i\theta} |+,+\rangle +|-,-\rangle -e^{i\theta}\left(|+,-\rangle
+|-,+\rangle \right)\right],\nonumber \hspace{85 pt}(12)\\
|\Phi_3\rangle &=&\frac{1}{\sqrt{2}}\left[-e^{2i\theta}
|+,+\rangle +|-,-\rangle\right],\nonumber\\
|\Phi_4\rangle &=&\frac{1}{\sqrt{2}}\left[|+,-\rangle - |-,+\rangle
\right],\nonumber
\end{eqnarray}
where $\theta=\varphi$ and $\theta=\varphi_1 + \varphi_2$ for the Hamiltonians (10) and (11) respetively.

Let us now take into consideration the two-atom one-mode  JCM. If atoms are initially  prepared in any of listed above eigenvalues
of semiclassical Hamiltonian (12) and field is initially prepared in one-mode coherent state of high intensity, the state
vector evolution may be described by the following asymptotic formulae:
\begin{eqnarray*}
|\Phi_1\rangle|\upsilon \rangle & \rightarrow & \frac12 \left\{
e^{-4igt}e^{ 2i\varphi}|+,+\rangle + |-,-\rangle + e^{-2igt}e^{-i\varphi}\left( |+,-\rangle +|-,+\rangle\right)
\right\}\times \nonumber\\
&\times&  \sum_{n=0}^{\infty} F_{n}|n\rangle
e^{-i 2 \Omega_n t},\nonumber \hspace{200 pt}(13)\\
|\Phi_2\rangle|\upsilon \rangle & \rightarrow  & \frac12 \left\{
e^{4igt}e^{ 2i\varphi}|+,+\rangle + |-,-\rangle -
e^{2igt}e^{-i\varphi}\left( |+,-\rangle - |-,+\rangle\right)
\right\}\times\nonumber\\ &\times&  \sum_{n=0}^{\infty} F_{n}|n\rangle
e^{i 2\Omega_{n} t},\nonumber \hspace{200 pt}(14)\\
|\Phi_3\rangle |\upsilon\rangle & \rightarrow & |\Phi_3\rangle
|\upsilon\rangle,  \,\,\,\,\,\,\,\,\,\ |\Phi_4\rangle
|\upsilon\rangle \rightarrow  |\Phi_4\rangle |\upsilon\rangle, \nonumber\hspace{138 pt}(15)\\
\end{eqnarray*}
where $\Omega_n \approx n$.

  It is clear from the expressions  (13)-(15) that the system state
vector can be factorized at any time of evolution. This means
that the system is in pure disentangled state at any time  under
the considered conditions. For the last two states
$|\Phi_3\rangle$ and $|\Phi_4\rangle$  the system state vector is
seen not to evolve at all. Of particular interest are the states
$|\Phi_1\rangle$ and $|\Phi_2\rangle$. There is no time
at which the atomic system prepared initially in the states
$|\Phi_1\rangle$ or $|\Phi_2\rangle$ is found in the same pure
state. But the atomic states appearing in Eqs. (12) and (13)
exactly coincide at time moments
$$
t_1 = (2k+1) T_{1R}/4, \eqno{(16)}$$
 where $k$ is integer and $T_{1R}$ is
one of Rabi oscillation revivals period.
 So, there are  series of
times for which  we can find $|\Phi_1\rangle$ and $|\Phi_2\rangle$ in the same pure state
$$\frac{1}{2} \left\{- e^{4i\varphi} |+,+\rangle + |-,-\rangle - \imath
e^{2i\varphi}\left( |+,-\rangle +|-,+\rangle\right) \right\}.$$
Thus, the disentanglement for atomic and field states occurs at
times $t_1$  only if the atomic system is initially
prepared in a linear superposition of two basis states
$|\Phi_1\rangle$ and $|\Phi_2\rangle$ such as $A$-state
$$|\Psi_A\rangle =\frac{1}{\sqrt{2}}\left(|+,-\rangle+|-,+\rangle\right) =
 e^{2i\varphi}\frac{1}{\sqrt{2}}\left(|\Phi_1\rangle - |\Phi_2\rangle\right)\eqno{(17)}$$
 and $B$-state
$$|\Psi_B\rangle=\frac{1}{\sqrt{2}}\left(e^{4\imath \varphi}|+,+\rangle+|-,-\rangle\right) =
\frac{1}{\sqrt{2}}\left(|\Phi_1\rangle
+|\Phi_2\rangle\right).\eqno{(18)}$$
 For these, the field state
at $t_1$  is a coherent superposition of macroscopically
distinct states which is usually called a 'Schr\"{o}dinger cat'.

Moreover, the field states in (13) and (14) exactly coincide (under condition $\bar n \gg 1$)
for times
$$t_2 = k\pi/2g=k T_{2R},\quad k =1,2, ... \eqno{(19)}$$

As a result, there exist two series of atom-field
disentanglement times for atoms prepared initially in the states
$|\Psi_A\rangle $ or $|\Psi_B\rangle$. In addition to this result
one can easily see from exact expression of  wave function (3) that atom-field
disentanglement takes place for all atomic initial states under
the conditions
$$
|\Omega_{n+2} - \Omega_{n+1}| t = 2 \pi k,\eqno{(20)}$$
$$
|\Omega_{n+1} - \Omega_{n}| t = 2 \pi k.\eqno{(21)}$$ For large
initial mean photon numbers $\bar n \gg 1$ equations (20) and (21)
are satisfying for times
$$t_3 = (\pi/g)k =T_{1R} k,\eqno{(22))}$$
where $k$ is integer. Thus, there is only one series of atom-field
disentanglement times for initial atomic states  distinct from $|\Psi_A\rangle $ or $|\Psi_B\rangle$.

 Mentioned results differ from these for both two-atom one-photon JCM \cite{Dung} and two-atom two-photon JCM \cite{bash2}. In the first one the
  disentanglement time  equals to a half of revival period for states (17) and (18), but for initial atomic states look
like $|+,+\rangle,\,|-,-\rangle,\,|+,-\rangle$  the state vector
for the whole system can not be presented as a product of its
subsystem state vectors at any time. For degenerate two-atom two-photon
 JCM  there are three series  of disentanglement times for $A$ and
$B$-states.

Let us now consider the two-atom two-mode JCM with intensity-dependent coupling. If atoms are initially prepared in any of listed above eigenvalues of
semiclassical Hamiltonian (12) and field is initially prepared in
two-mode coherent state of high intensity, the state vector evolution
may be described by the following asymptotic formulae:
\begin{eqnarray}
|\Phi_1\rangle|\upsilon_1,\upsilon_2\rangle&\rightarrow&\frac12\sum_{n_1,n_2=0}^{\infty}
F_{n_1}F_{n_2}|n_1\rangle|n_2\rangle
e^{-i \Omega_2(n_1,n_2)t}\times\nonumber\\
&\times&\left\{ e^{2i\left(\varphi_1+\varphi_2\right)}e^{-8igt
(n_1+n_2) }|+,+\rangle + |-,-\rangle+\right.\nonumber\\
&+&\left.e^{i\left(\varphi_1+\varphi_2\right)}e^{-4igt(n_1+n_2)
}\left(
|+,-\rangle +|-,+\rangle\right) \right\},\nonumber\\
|\Phi_2\rangle|\upsilon_1,\upsilon_2\rangle&\rightarrow&\frac12\sum_{n_1,n_2=0}^{\infty}
F_{n_1}F_{n_2}|n_1\rangle|n_2\rangle
e^{i\Omega_2(n_1,n_2)t}\times \nonumber\\
&\times&\left\{
e^{2i\left(\varphi_1+\varphi_2\right)}e^{8igt(n_1+n_2)}|+,+\rangle
+ |-,-\rangle - \right. \nonumber\\
&-&\left.
e^{i\left(\varphi_1+\varphi_2\right)}e^{4igt(n_1+n_2)}\left(
|+,-\rangle +|-,+\rangle\right) \right\},\nonumber\\
|\Phi_3\rangle |\upsilon_1,\upsilon_2\rangle &\rightarrow&
|\Phi_3\rangle |\upsilon_1,\upsilon_2\rangle, \,\,\,\,\,\,\,\,\,\
|\Phi_4\rangle |\upsilon_1,\upsilon_2\rangle \rightarrow
|\Phi_4\rangle |\upsilon_1,\upsilon_2\rangle .\nonumber
\end{eqnarray}
\noindent
There are  times at which the atomic system prepared
initially in the states $|\Phi_1\rangle$ and $|\Phi_2\rangle$ is found in the same pure atomic
state. These times  are
$$t_{4}=\frac{\pi m}{2g},\quad m =0,1,2\dots \eqno{(23)}.$$
Unlike the previous model the times (23) differ from the periods of Rabi oscillation revivals (8) and (9).
 The times $t_4$ do not depend on mean photon numbers in the cavity modes. For  states
$|\Phi_3\rangle$ and $|\Phi_4\rangle$  the system state vector is
seen not to evolve at all.

Thus, the disentanglement for atomic and field states occurs only if
the atomic system is initially prepared in a linear superposition of
two basis states $|\Phi_1\rangle$ and $|\Phi_2\rangle$ such as
$|\Phi_1\rangle$ and $|\Phi_2\rangle$:
$$|\Psi'_B\rangle=\frac{1}{\sqrt{2}}\left(|+,-\rangle+|-,+\rangle\right) =
 e^{i\left(\varphi_1+\varphi_2\right)}\frac{1}{\sqrt{2}}\left(|\Phi_1\rangle - |\Phi_2\rangle\right) $$
 and
$$|\Psi'_B\rangle=\frac{1}{\sqrt{2}}\left(e^{2\imath(\varphi_1+\varphi_2)}|+,+\rangle+|-,-\rangle\right) =
\frac{1}{\sqrt{2}}\left(|\Phi_1\rangle +|\Phi_2\rangle\right).$$

One can also easily see that field states in expressions
 $|\Phi_1\rangle|\upsilon_1,\upsilon_2\rangle$  and $|\Phi_2\rangle|\upsilon_1,\upsilon_2\rangle$ are  coincide  (under the condition $\bar n_1,\bar n_2 \gg 1$) for times  $t_5$  satisfying the formula
$$\Omega_2(n_1,n_2) t_5 =  \pi m ,\quad m =1,2, .... $$
From these conditions one can obtain that $t_5=t_4$. As a result, there exists only one series of atom-field disentanglement times for atoms prepared initially in the states  $|\Psi'_A\rangle $ and $|\Psi'_B\rangle$.
 The deriving  of atom-field disentanglement conditions for arbitrary initial atomic pure state is aim of our following paper.

Thus we have derive the times of atom-field disentanglement for initial atomic states $|\Psi'_A\rangle $ and $|\Psi'_B\rangle$.    This result differs from that  for two-atom  one-mode JCM  with intensity-dependent coupling. In the previous model we have two series of atom-field disentanglement times for  $A$ and $B$ initial atomic states.

\section{The dynamics of the reduced atomic entropy for various
initial atomic and field states}
\label{sec:3}

Analytical conclusions about the system state vector dynamics and
atom-field entanglement can be verified through  linear entropy
examining. A linear entropy of reduced atomic (or field) density matrix
can serve for entanglement degree evaluation of the systems
consisting of two subsystems and being prepared in a pure state.
The linear entropy of reduced atomic density matrix for
considered systems has the following form:
   $$
 S=1-Tr\left(\rho_{AT}^{2}\right)\eqno{(24)}$$
 where $\rho_{AT} = Tr_F (\mid \Psi\rangle \langle \Psi \mid)$.  The case when $S=0$  corresponds
to completely
disentangled atomic and field states, and the case when a linear entropy equals to $\>\>3/4$   corresponds to maximum entanglement
degree. For numerical calculations of the linear atomic entropy (24) one can  use the fact that the Poissonian distribution for coherent states has the width which is proportional to
$\sqrt{\overline{n}}$. So, we have  restricted ourselves to the finite sums.

 \begin{figure}
\begin{tabular}{c}
\mbox{(a)} \\
\includegraphics[width=0.75\textwidth]{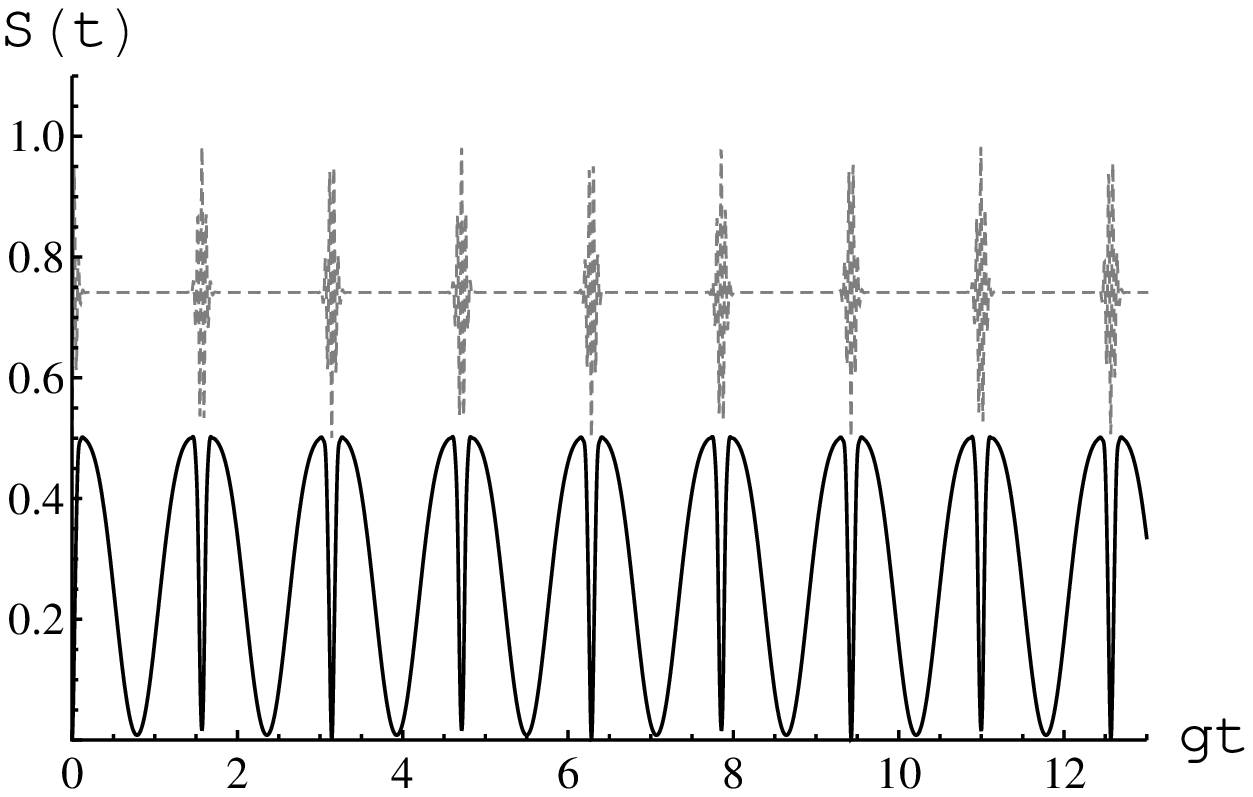} \\
\mbox{(b)} \\
\includegraphics[width=0.75\textwidth]{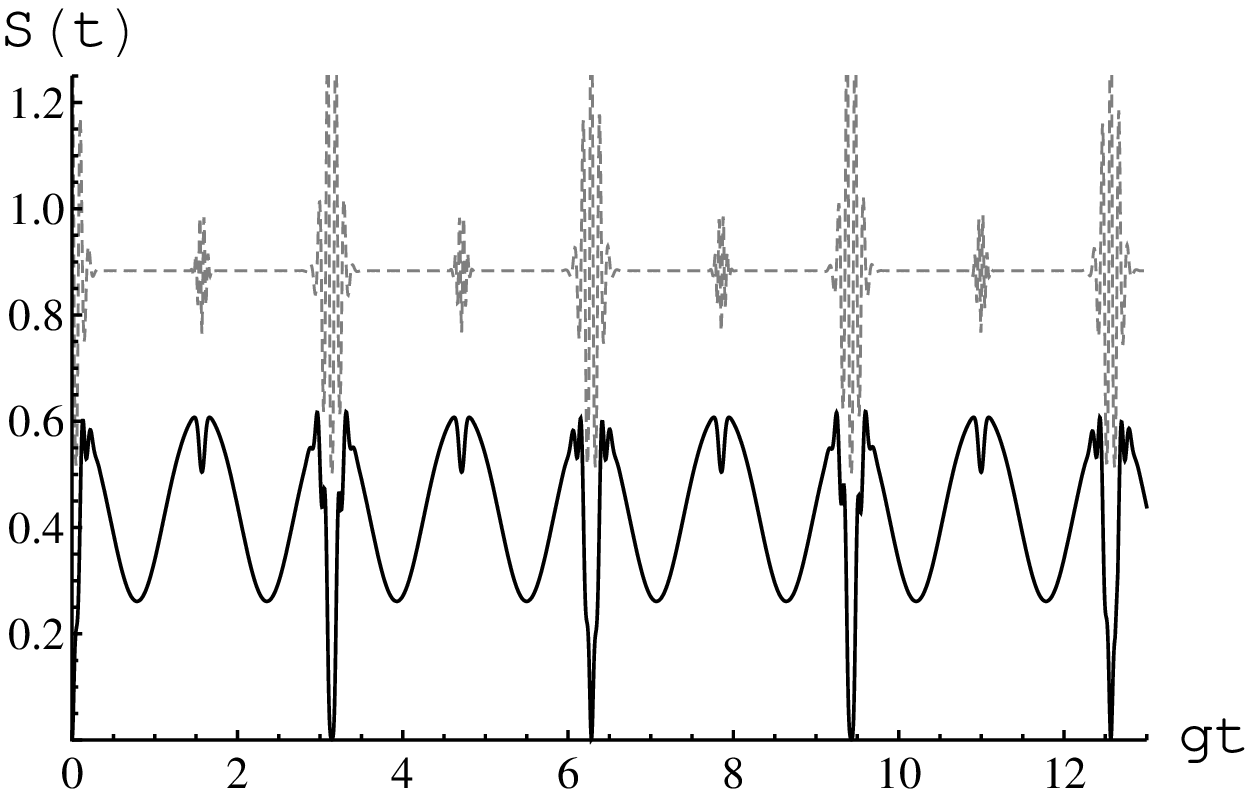} \\
\end{tabular}
\vspace*{8pt}
\caption{Time evolution of linear entropy for atomic initial states:
(a) $\sqrt{1/2}(|+,-\rangle + |-,+\rangle)$ and (b) $|+, +\rangle$. The initial mean photon number  $\overline{n} =30$ and phase of coherent state $\varphi=0$. Gray curves denote the possibility to find both atoms in excited stated $W_{++}+0.5$.}
\label{fig:1}
\end{figure}

For the first model the dynamics of atomic linear entropy
$S$ is presented in Fig. 1 for
different initial atomic states and a coherent one-mode field of
high intensity. We draw both values of linear entropy
$S$  (black curves) and atomic possibilities to find both
atoms in excited state (gray curves) in the figures.  Fig. 1a
demonstrates the time behavior of linear atomic entropy for the
case when the atomic subsystem is prepared initially in state
$A$ (or $B$).
 One can easily see from Fig.1a that there are two series of disentanglement times which
  exactly coincide with values predicted in Section II (see formulae (16) and (19)).
    The result
confirms well the conclusions made on the basis of the analysis
of the state vector asymptotic dynamics. As for the states
$|+,+\rangle$ (or $|-,-\rangle, |+,-\rangle$ and $|-,+\rangle$) it
can be clearly seen that the system evolves into entangled state
and  revive into its  disentangled one only for times described exactly by formula (22)( see Fig. 1b).

So, one can see that all the results obtained by the linear
entropy numerical calculations are in good accordance with the
analytical expressions for wave function  made in the previous
section.

\begin{figure}
\begin{tabular}{c}
  \mbox{(a)}\\
\includegraphics[width=0.75\textwidth]{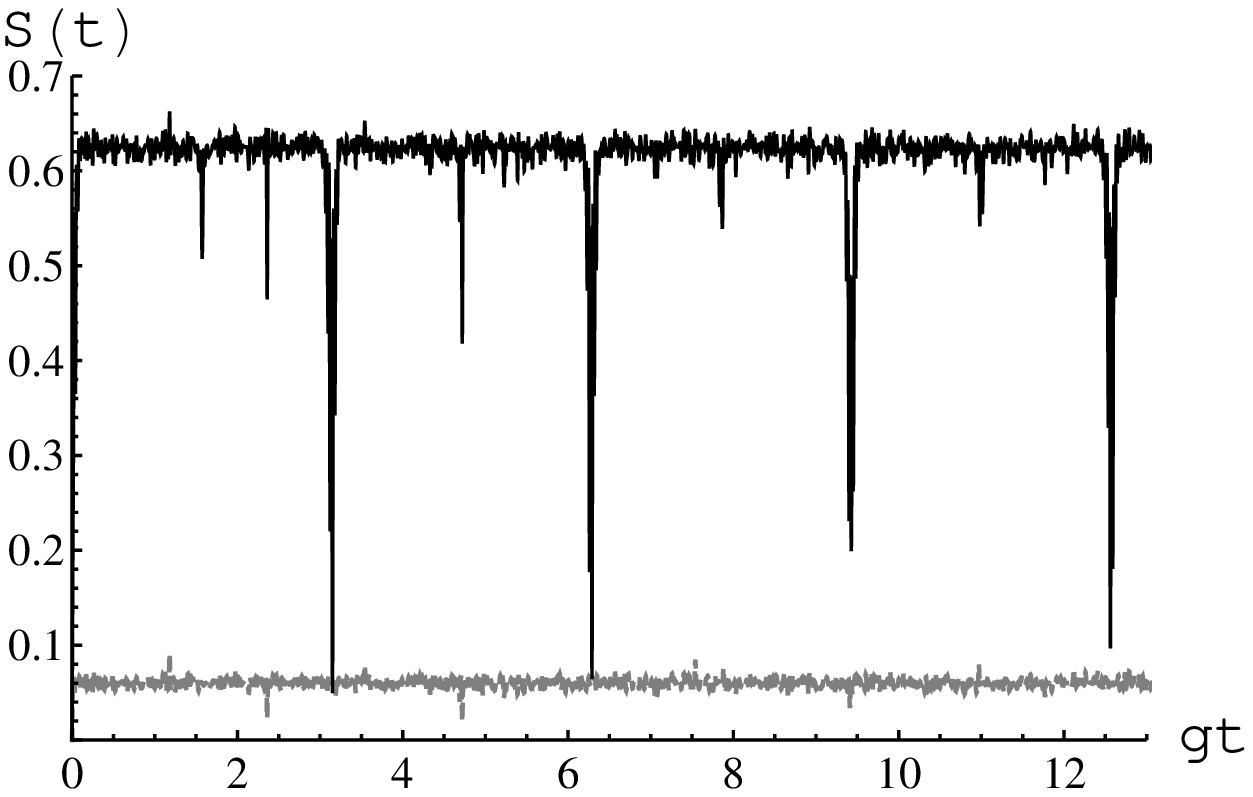} \\
 \mbox{(b)}\\
\includegraphics[width=0.75\textwidth]{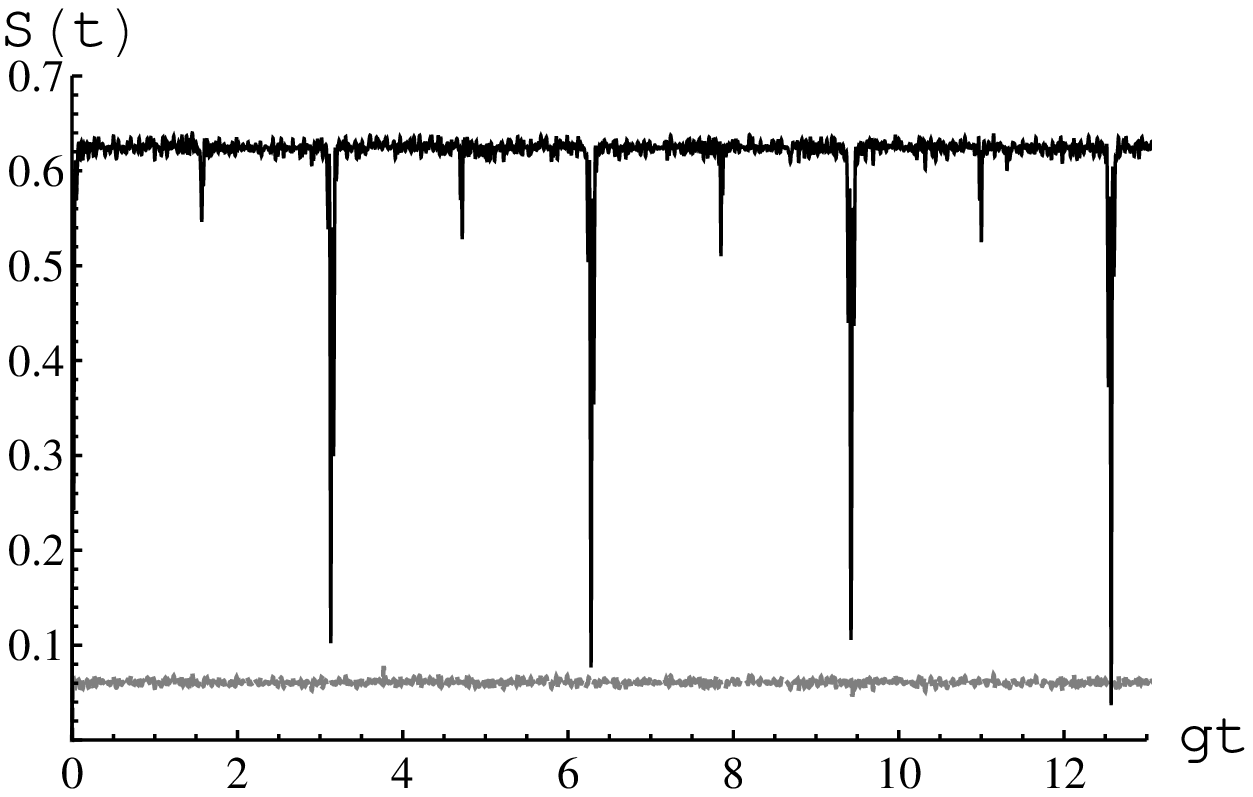} \\
\end{tabular}
\vspace*{6pt}
\caption{Time evolution of linear entropy for initial atomic state   $(1/\sqrt{2})|+,-\rangle + |-,+\rangle$. The mean photon numbers  $\overline{n}_1 =\overline{n}_2= 50$ (Fig 1a) and $\overline{n}_1 =50, \overline{n}_2= 150$ for (Fig 1b). The phases of the coherent states  $\varphi_1=\varphi_2=0$. Gray curves denote the possibilities to find both atoms in exited states $|+,+\rangle$.}
\label{fig:2}
\end{figure}
\begin{figure}
\begin{tabular}{c}
  \mbox{(a)}\\
\includegraphics[width=0.75\textwidth]{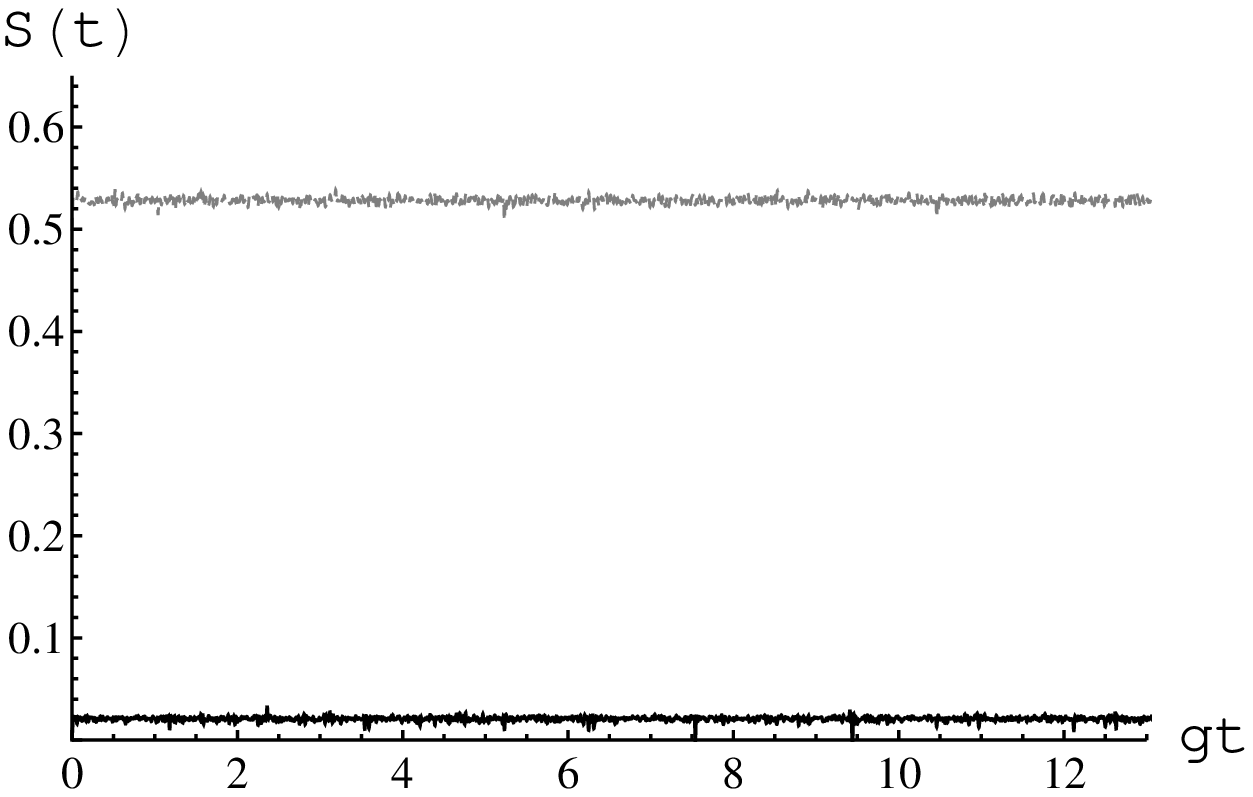} \\
 \mbox{(b)}\\
\includegraphics[width=0.75\textwidth]{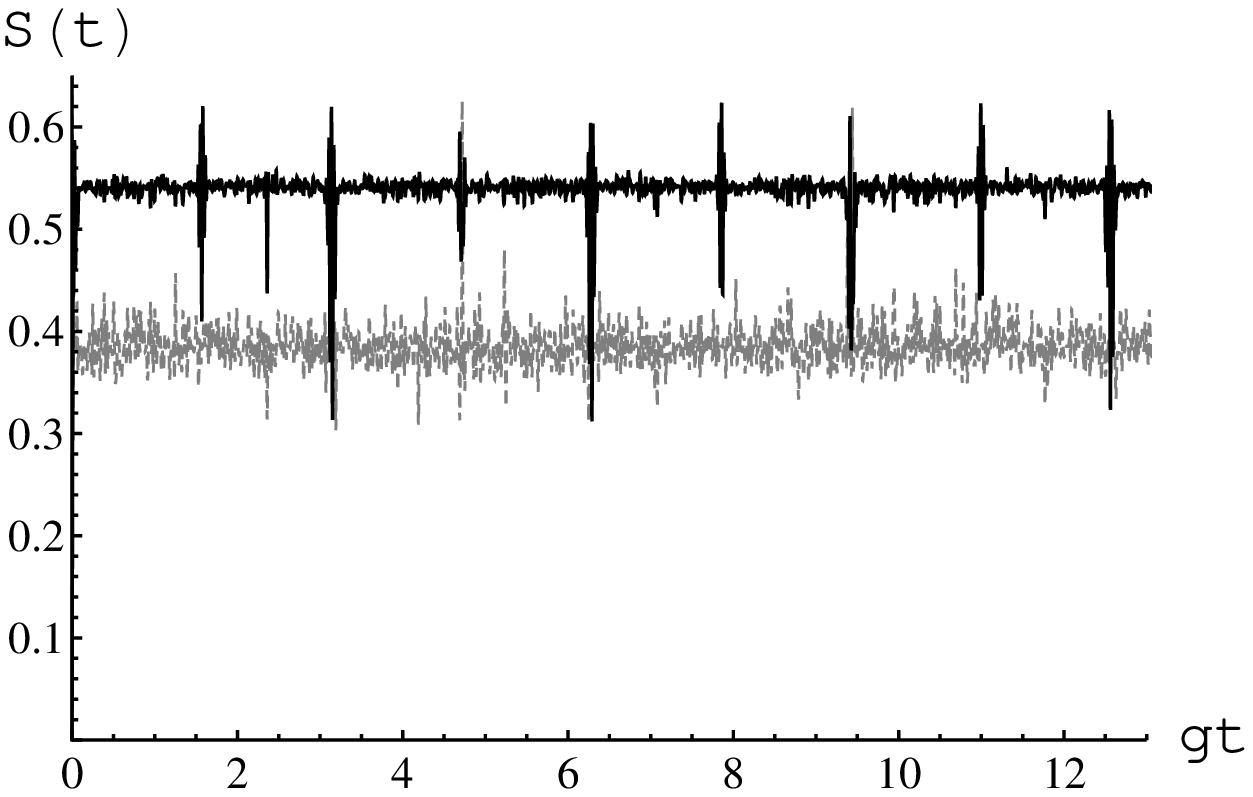} \\
\end{tabular}
\vspace*{6pt}
\caption{Time evolution of linear entropy for initial atomic states:
   $|\Phi_3\rangle$ (a) and  $|+, +\rangle$ (b). The mean photon numbers  $\overline{n}_1 =\overline{n}_2= 50$. Gray curves denote the possibilities to find both atoms in exited states $|+,+\rangle$.}
\label{fig:3}
\end{figure}
For the second model the dynamics of atomic linear entropy  is presented in Fig. 2 and 3
 for different initial atomic states and a coherent two-mode
field of high intensity. Fig 2 shows the time behavior of linear atomic entropy for initial atomic state $|\Psi_A\rangle$ and  two choices of the mean photon numbers in two-mode coherent cavity field: $\bar {n}_1 =  \bar {n}_2 = 50$ (Fig. 2a)
and $\bar {n_1} = 50,   \bar {n_2} = 150$ (Fig. 2b).
 We draw both values of linear entropy
$S$  (black curves) and atomic possibilities to find both
atoms in excited state (gray curves) in the figures.
One can easily see from Fig. 2 that there one
series of disentanglement times which don't depend on mean photon numbers and exactly coincide with values (23) predicted in Section 2 for states $A$  and  $B$. We have obtained
the atom-field entanglement which is periodically changed with completely full revival into pure disentangled state.
Fig 2 shows the time behavior of linear atomic entropy for initial atomic state $|+,+\rangle$ and   $\Phi_3$ for $\bar {n}_1 =  \bar {n}_1 = 50$. One can easily sea from Fig. 3a that for initial atomic state
$|\Phi_3\rangle$   the system is really in a pure initial state disentangled state at any time.
  The analysis of analytical and numerical calculations of linear entropy for state $|+,+\rangle$ presented in Fig. 3b is too difficult. These problem will be solved in our following paper.
So, one can see that all the results obtained by the linear entropy numerical calculations are in good accordance with the analytical
expressions for wave function made in the previous section  for states $A$ and $B$.

\section{Conclusions}
\label{sec:4}

Two-atom one-mode and two-mode JCM with intensity-dependent coupling  are considered in the paper.
Atom-field entanglement is shown to occur in both models, the
entanglement degree is estimated on the basis of analysis of wave-function behavior and a linear entropy
criterion. The disentanglement is found to appear in the  models
for some initial atomic states and large coherent field inputs with different values
of the mode intensities relation. We also estimate the periods
of disentanglement for the models.

\begin{acknowledgements}
Author thanks M.S. Rusakova and E.Yu. Sochkova for help in calculations.
\end{acknowledgements}

\end{document}